\documentclass[superscriptaddress,aps,prx,onecolumn,floatfix,10pt]{revtex4-2}

\usepackage{times}
\usepackage{color,soul}

\usepackage{graphicx}

\usepackage{hyperref}
\usepackage{amsmath}

\newenvironment{sciabstract}{%
\begin{quote} \bf}
{\end{quote}}

\usepackage{verbatim}

\newcommand{\detailtexcount}[1]{%
  \immediate\write18{texcount -merge -sum -q #1.tex output.bbl > #1.wcdetail }%
  \verbatiminput{#1.wcdetail}%
}

\begin{document}

\detailtexcount{main}

\newpage
\title{Observation of Superconducting Solitons by Terahertz-Light-Driven Persistent Pseudo-Spin Coherence}




\author{M. Mootz}
 \thanks{These authors contributed equally to this work.}
 \affiliation{Ames National Laboratory, Ames, IA 50011, USA}
\author{C. Vaswani}
 \thanks{These authors contributed equally to this work.}
 \affiliation{Ames National Laboratory, Ames, IA 50011, USA}
 \affiliation{Department of Physics and Astronomy, Iowa State University, Ames, IA 50011, USA}
\author{C. Huang}
 \affiliation{Ames National Laboratory, Ames, IA 50011, USA}
\affiliation{Department of Physics and Astronomy, Iowa State University, Ames, IA 50011, USA}
 \author{K.~J.~Lee}
\affiliation{Department of Materials Science and Engineering, University of Wisconsin-Madison, Madison, WI 53706, USA}
\author{A. Khatri}
\affiliation{Ames National Laboratory, Ames, IA 50011, USA}
\affiliation{Department of Physics and Astronomy, Iowa State University, Ames, IA 50011, USA}
\author{P.~Mandal}
 \affiliation{Department of Materials Science and Engineering, University of Wisconsin-Madison, Madison, WI 53706, USA}
\author{J.~H.~Kang}
 \affiliation{Department of Materials Science and Engineering, University of Wisconsin-Madison, Madison, WI 53706, USA}
\author{L. Luo}
 \affiliation{Ames National Laboratory, Ames, IA 50011, USA}
\author{I.~E.~Perakis}
 \affiliation{Department of Physics, University of Alabama at Birmingham, Birmingham, AL 35294-1170, USA}
\author{C.~B.~Eom}
 \affiliation{Department of Materials Science and Engineering, University of Wisconsin-Madison, Madison, WI 53706, USA}
\author{J. Wang}
 \email{To whom correspondence should be addressed: jgwang@iastate.edu, jgwang@ameslab.gov}
 \affiliation{Ames National Laboratory, Ames, IA 50011, USA}
 \affiliation{Department of Physics and Astronomy, Iowa State University, Ames, IA 50011, USA}

\date{\today}

\maketitle
\baselineskip20pt


	
\begin{sciabstract}
Overcoming the decoherence bottleneck remains a central challenge for advancing coherent superconducting quantum device and information technologies.
Solitons---non-dispersive wave packets stabilized by the collective synchronization of  quantum excitations---offer a robust pathway to mitigating dephasing, yet their realization in superconductors has remained experimentally elusive. 
Here, we report the observation of a driven soliton state in epitaxial thin films of an iron-based superconductor (Co-doped BaFe$_2$As$_2$), induced by intense, multi-cycle terahertz (THz) periodic driving.
The dynamical transition to this soliton state is marked by the emergence of Floquet-like spectral sidebands that exhibit a strongly nonlinear dependence on THz laser field strength and a  resonant enhancement with temperature. 
Quantum kinetic simulations corroborate these observations, allowing us to underpin the emergence of synchronized Anderson pseudo-spin oscillations---analogous to Dicke superradiance---mediated by persistent order parameter oscillations. In this coherently driven state, the observed sidebands result from difference-frequency mixing  between the THz drive and persistent soliton dynamics.
These findings establish a robust framework for coherently driving and controlling superconducting soliton time-crystal-like phases using low dissipation, time-periodic THz fields, enabling prospects for THz-speed quantum gate operations, long-lived quantum memory, and robust quantum sensing based on enhanced macroscopic pseudo-spin coherence.
\end{sciabstract}

\section{Introduction}
Non-equilibrium superconducting (SC) condensates display a wide range of quantum states distinguished by their time-dependent order parameters and exotic collective modes~\cite{Blumberg,Fausti2011,Matsunaga:2013,matsunaga2014,caval,Mitrano2016,Matsunaga2017,Yang2018b,yang2018,yang2019lightwave,Giorgianni2019,vaswani2019discovery,Chu2020,Budden2021,Shimano2021,Buzzi2021,hybrid-higgs,Luo2023, nanlin}. A particularly intriguing theoretical prediction is the occurrence of non-equilibrium superconducting states  with soliton behavior, characterized by a time-dependent order parameter  displaying persistent, undamped oscillations at THz frequencies~\cite{Yuzbashyan2008}. In synthetic quantum matter with ultracold atomic gases, such soliton states have been realized  through quantum quench of the interactions~\cite{Yuzbashyan2006}, or by using cavity quantum electrodynamics~\cite{Young2024}.
Such states are also of technological interest, particularly for the development of quantum computing and memory systems with both high coherence and THz speed.
However, dynamical superconducting soliton states have, so far, been realized only at sub-GHz frequencies in cold atomic gases, with no soliton signatures reported in solid state superconductors yet.

Recent advancements in ultrafast THz coherent spectroscopy now  enable the implementation of a THz quantum quench with minimal heating in superconducting quantum materials~\cite{Yang2023}. They  also allow coherent control of the post-quench   quantum dynamics in a pre-thermalized state, as well as during time-periodic driving by multi-cycle 
THz fields. 
To date, THz laser-driven quantum quenches of superconductor order parameters have led to diverse outcomes, including Landau-damped Higgs oscillations~\cite{Matsunaga:2013,matsunaga2014,hybrid-higgs,Luo2023, echo}, overdamped dynamics resembling a gapless quasiparticle (QP) metastable phase~\cite{yang2018}, and high harmonic pseudo-spin precession exhibiting multiple frequencies both allowed and forbidden by particle-hole symmetry~\cite{yang2019lightwave,vaswani2019discovery}.  
However, soliton driven  states remain elusive in superconductors. 

\begin{figure}[ht!]
    \centering
    \includegraphics[width=0.8\textwidth]{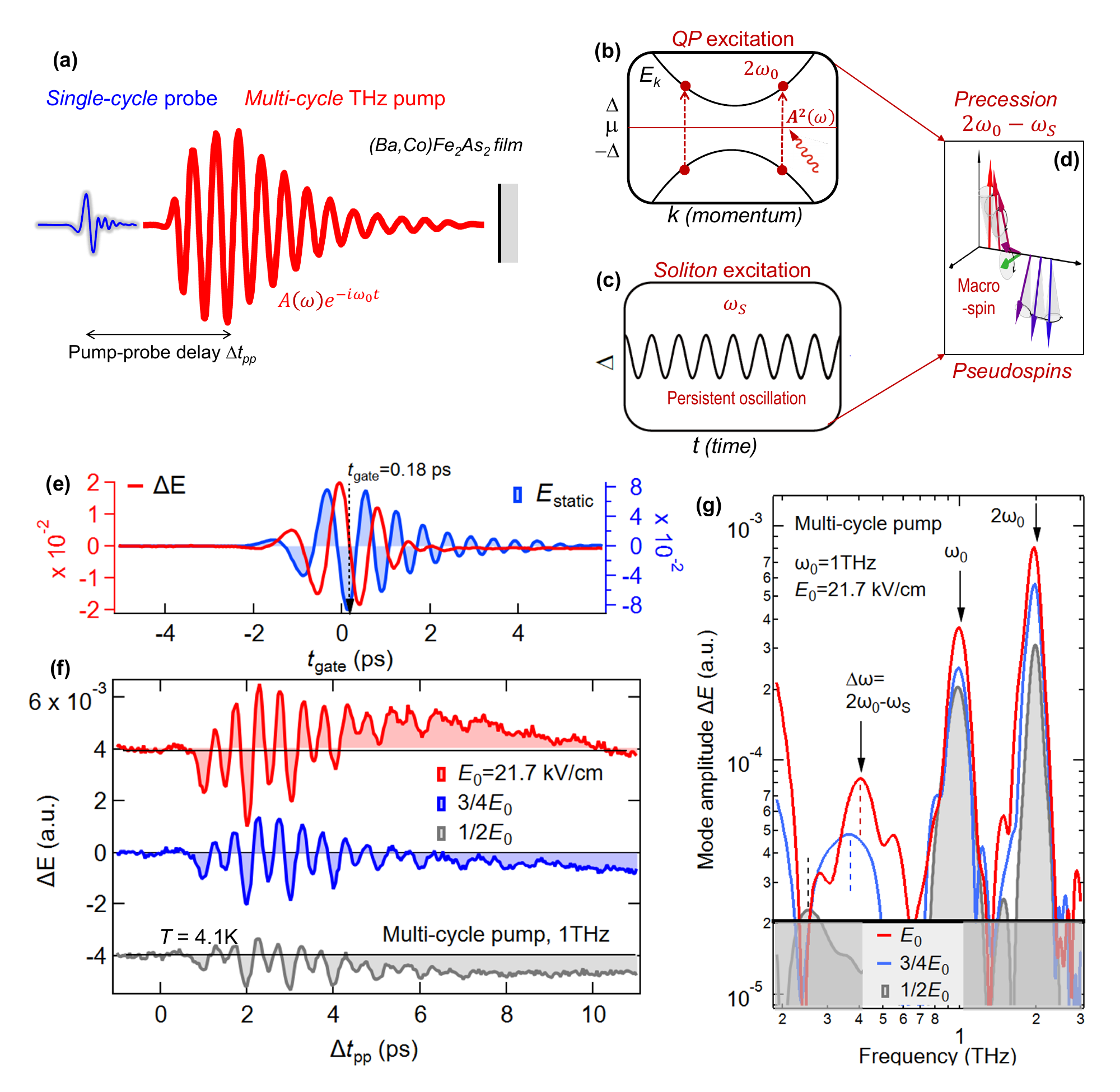}
    \caption{Experimental observation of THz-driven pseudo-spin soliton dynamics in a multiband superconductor. 
(a) Schematic of the pump--probe experimental setup and pulse sequence. The Ba(Fe$_{1-x}$Co$_{x}$)$_2$As$_2$ ($x$ = 0.08) epitaxy film is excited by a multi-cycle narrowband THz pump pulse (red line), followed by a broadband single-cycle THz probe pulse (blue line) used to detect the induced non-equilibrium state. 
(b) Schematic illustration of strong THz driving above the superconducting gap ($2\omega_0 \gtrsim 2\Delta_\mathrm{SC}$): intense multi-cycle THz fields induce coherent quasiparticle (QP) population inversion within the continuum, producing pseudo-spin Rabi oscillations that access the entire Bloch sphere.
(c) Following these Rabi oscillations, the pseudo-spin ensemble evolves into a synchronized state: individual pseudo-spins with initially distributed phases lock together during the latter part of the THz pulse, mediated by nonlinear coupling via the oscillating superconducting order parameter. This synchronized pseudo-spin collective state corresponds to a soliton state characterized by an order parameter oscillating at  frequency $\omega_\mathrm{S}$ without damping. 
(d) Illustration of difference-frequency pseudo-spin precession arising from nonlinear mixing between the collective soliton oscillation at $\omega_\mathrm{S}$ and the high-frequency pseudo-spin precession at $2\omega_0$. In the soliton state, many pseudo-spins synchronize into a collective "giant" pseudo-spin precessing at $\omega_\mathrm{S}$, which coherently interacts with the $2\omega_0$ precession to generate a new mode at $2\omega_0 - \omega_\mathrm{S}$. 
(e) Measured differential transmitted THz field $\Delta E(t_\mathrm{gate})$ at $T=4.1$~K and pump field $E_\mathrm{THz}=21.7$~kV/cm with fixed pump--probe delay $\Delta t_\mathrm{pp}=3$~ps, showing a coherent oscillatory response. 
The blue line represents the transmitted THz probe pulse, $E_\mathrm{static}$, which was measured concurrently during the experiment.
(f) Dynamics of $\Delta E(\Delta t_\mathrm{pp})$ at fixed gate time $t_\mathrm{gate}=0.18$~ps (dashed line in (e)) for three pump field strengths: $E_0=21.7$~kV/cm (red), $3/4\,E_0=16.3$~kV/cm (blue), and $1/2\,E_0=10.9$~kV/cm (gray). Distinct quantum beating and amplitude modulation patterns appear and become more pronounced at higher fields.
(g) Fourier spectra of the oscillatory signals in (f) showing not only the fundamental $\omega_0$ and second harmonic $2\omega_0$ peaks but also a prominent low-frequency sideband $\Delta \omega$ (0.2--0.4~THz) far below the superconducting gap. The emergence and strong nonlinear growth of this sideband with increasing field provide first experimental evidence for the THz-induced pseudo-spin soliton state.
}
    \label{fig1:main}  
\end{figure}

Recent studies of coherent ultrafast nonlinear dynamics have identified  unconventional Fe-based superconductors (FeSCs) as a distinctive model system. Excitation with single-cycle THz pulses has been shown to produce a quantum quench of the SC order parameter $\Delta$ resulting in distinct quantum dynamics as compared to conventional single-band 
SCs, attributed to the unique Coulomb-coupled electron ($e$) and hole ($h$) bands of these materials. In particular, unlike in other multi-band SCs, the condensates in  different bands  are coupled by a robust interband $e$--$h$ interaction $U$ that well exceeds  the intraband interaction $V$. Such strongly coupled multi-band condensates exhibit unique collective modes, including hybrid Higgs~\cite{hybrid-higgs} and phase-amplitude modes~\cite{Luo2023} absent for small $U$. 
Unlike previous studies employing broadband, single-cycle excitation of these superconductors, here we demonstrate that a soliton SC state emerges while this system is time-periodically driven by multi-cycle THz electric field pulses, as illustrated in Fig.~\ref{fig1:main}(a) and detailed below.

\section{Experimental Design}
Figures~\ref{fig1:main}(b)--\ref{fig1:main}(d) illustrate the physical scenario  motivating our experiment. The conventional model for describing nonlinear driving of moderately clean SC systems by multi-cycle THz pulses with central frequency $\omega_0$ involves the precession of Anderson pseudo-spins  driven  by even powers $\mathcal{O}(\mathbf{A}^{2n})$ of the laser vector electromagnetic potential $\mathbf{A}(t)$,  described by BCS-like Hamiltonians introduced  in previous works~\cite{matsunaga2014,Aoki2017,Forster2017}. For relatively weak {\em resonant} excitation of the SC system, i.~e. $2\omega_0=2\Delta_\mathrm{SC}$ where $2\Delta_\mathrm{SC}$ is the SC gap, the solution of these BCS-like Hamiltonians reveals a non-equilibrium steady-state with finite order parameter  following  the decay of  Higgs collective mode oscillations, whose frequency $2\Delta_\mathrm{SC}$ is close to the laser second harmonic frequency $2\omega_0$. On the other hand, strong THz excitation {\em above the SC gap}  yields an inverted coherent  population of QPs  within the continuum (Fig.~\ref{fig1:main}(b)), assisted by Rabi oscillations.
Analogous to the unstable equilibrium state of an inverted pendulum, such
 coherent QP population inversion creates an unstable initial state whose time evolution can lead to the formation of a   soliton SC state characterized by an order parameter displaying undamped time oscillations (Fig.~\ref{fig1:main}(c)). 

More specifically,  strong excitation with a multi-cycle THz pulse at or above the SC gap results in pseudo-spin Rabi oscillations within the QP continuum. Rabi oscillations allow access to the entire Bloch sphere with increasing  field, which can  invert  the coherent QP population  of  the pseudo-spin texture in the vicinity of $2\omega_0$. Such coherent QP population inversion serves as a nonlinear initial condition for soliton collective effects~\cite{Balseiro}. As depicted in Fig.~\ref{fig1:main}(c), the time-dependent pseudo-spin distribution across a range of momenta can  synchronize after few THz oscillation cycles, in a way analogous to the development of Dicke superradiance~\cite{Dicke,Yuzbashyan2005,Yuzbashyan2015}. This synchronization  of many QP coherent excitations  with locked relative phases  arises from their nonlinear coupling  mediated self-consistently by  the oscillating superconducting order parameter. The properties of the  Floquet-like state driven  by the superconducting order parameter oscillations  at the  soliton frequency $\omega_\mathrm{S}$ can be described in terms of the precession of a collective ``giant'' pseudo-spin, rather than by the precession of individual pseudo-spins with a distribution of frequencies determined by the band dispersion (Fig.~\ref{fig1:main}(b)).

Using the experimental geometry shown in Fig.~\ref{fig1:main}(a), our quantitative simulations of multi-cycle pump and single-cycle probe dynamics predict the nonlinear emergence of distinct, experimentally accessible satellite spectral features under strong Coulomb coupling between electron and hole bands (Appendix~A).
 These satellite  spectral peaks, centered   at  frequencies $\Delta \omega \sim 2\omega_0- \omega_\mathrm{S}$ well below the superconducting energy gap $2\Delta$, 
are generated during the latter part of the multi-cycle laser pulse via  difference-frequency nonlinear processes. These processes arise from the interaction between the  soliton collective mode  oscillating at frequency $\omega_\mathrm{S}$ and the pseudo-spin coherence  excited by the laser  at frequencies  $\sim 2\omega_0$ (Fig.~\ref{fig1:main}(d)). Here we report the experimental observation of such spectral satellites arising for multi-cycle narrowband THz excitation  of an optimally-doped Ba(Fe$_{1-x}$Co$_{x}$)$_2$As$_2$ superconductor. 
The robust, low-energy $\Delta \omega$ collective oscillations intensify when the pseudo-spin soliton sideband frequency resonates with the $2\Delta$ gap, giving rise to a distinctive dependence on temperature and THz photon frequency---hallmarks for the emergence of collective mode sidebands at sub---harmonic frequencies well below $2\Delta_\mathrm{SC}$. 
Furthermore, our experimental results, comparing FeSCs with single-band BCS superconductors, are consistent with quantum kinetic simulations predicting that strong interband interactions ($U$) within a multiband condensate facilitate the formation of a driven Anderson pseudo-spin soliton state characterized by a low-energy Floquet-like sideband.

\section{Results and discussion}
We measure an epitaxial Ba(Fe$_{1-x}$Co$_{x}$)$_2$As$_2$ ($x = 0.08$) film of 60~nm thickness, grown on a 40~nm thick SrTiO$_3$ buffered (001)-oriented (La;Sr)(Al;Ta)O$_3$ (LSAT) single crystal substrate with $T_\mathrm{c}\sim$ 23.4~K (Sec. 1, Appendix~B). The epitaxial and good crystalline quality of our sample was characterized by four-circle X-ray diffraction, together with extensively chemical, optical, structural, and electrical characterizations. A zero-field-cooled magnetization using superconducting quantum interference device (SQUID) magnetometer measurements shows a diamagnetic signal. We perform a THz pump and THz probe experiment,  measuring the differential THz transmission $\Delta E$ as a function of pump--probe time delay $\Delta t_\mathrm{pp}$. The SC state of our  Ba(Fe$_{0.92}$Co$_{0.08}$)$_2$As$_2$ film is excited by a multi-cycle, narrowband THz pump pulse with frequency $\omega_\mathrm{0}=1$~THz (4.1~meV) or 0.5~THz (2.1~meV). The 0.5~THz (1~THz) pulse with $2\omega_0=4.1$~meV ($2\omega_0=8.2$~meV) excites below (above) the dominant pair-breaking gap, $2\Delta_\mathrm{SC}\simeq 6.8$~meV at 4.1~K (Sec. 2, Appendix~B). The pump electric field waveforms with multiple oscillation cycles (Sec. 3, Appendix~B). The probe pulse is broadband and nearly single-cycle (blue line, Fig.~\ref{fig1:main}(a)). Such broadband probe is used to detect the non-equilibrium state subsequent to the multi-cycle narrowband pump excitation. Characterization of this pump-driven state is achieved through the measurement of the induced THz probe field transmission across the excited sample, denoted as $\Delta E(t_\mathrm{gate}, \Delta t_\mathrm{pp})$, wherein both the gating pulse time $t_\mathrm{gate}$ and the pump--probe time $\Delta t_\mathrm{pp}$ are systematically scanned. An example is presented by the red trace in Fig.~\ref{fig1:main}(e), showcasing raw data of $\Delta E(t_\mathrm{gate})$ obtained at 4.1~K and $\Delta t_\mathrm{pp}=$3~ps, with the transmitted THz probe pulse, $E_\mathrm{static}$ (blue trace), which was measured concurrently during the experiment. 

The coherent oscillations observed for our Ba(Fe$_{0.92}$Co$_{0.08}$)$_2$As$_2$ film at 4.1~K are plotted in Fig.~\ref{fig1:main}(f). We compare the responses for three different  driving electric fields, namely $E_\mathrm{THz}=E_0=21.7$~kV/cm (red line), $3/4 E_0=16.3$~kV/cm (blue line), and $1/2 E_0=10.9$~kV/cm (gray line). All three fields are centered  at the same frequency $\omega_0=1$~THz. The gate time $t_\mathrm{gate}=0.18$~ps is kept fixed (dashed line in Fig.~\ref{fig1:main}(e)). The dynamics of $\Delta E(\Delta t_\mathrm{pp})$ reveals a distinct quantum beating superimposed on a gradual amplitude modulation. Figure~\ref{fig1:main}(g) displays the spectra of those oscillatory $\Delta E$ signals, obtained through Fourier transformation of the raw temporal dynamics. In addition to the spectral peak at the fundamental frequency $\omega_0$, which partially arises from a THz-field-accelerated supercurrent~\cite{Mootz2022}, these spectra exhibit two additional  prominent peaks: (i) The conventional quadratic nonlinearity peak centered at  the pseudo-spin precession frequency 2$\omega_0$; (ii) A low-frequency distinctive  peak, which emerges well below $\omega_0$ and the energy gap $2\Delta_\mathrm{SC}$, at frequency  $\Delta \omega \sim$ 0.2--0.4~THz. The latter new peak develops  nonlinearly with increasing  electric field and becomes pronounced under high THz driving: compare $E_0$ (red line), $3/4 E_0$ (blue line), and $1/2 E_0$ (gray line) results in Fig.~\ref{fig1:main}(g). We will show next that this new low-frequency sideband peak  is the hallmark of a Floquet soliton state. 
Note that a background ``overshoot" appears towards zero frequency, arising from the slowly varying amplitude envelope in Fig.~\ref{fig1:main}(f). Under this excitation condition and in the low-temperature regime, the background shows minimal spectral overlap with the soliton sideband at frequency $\Delta \omega$.

\begin{figure}[ht!]
    \centering
    \includegraphics[width=0.8\textwidth]{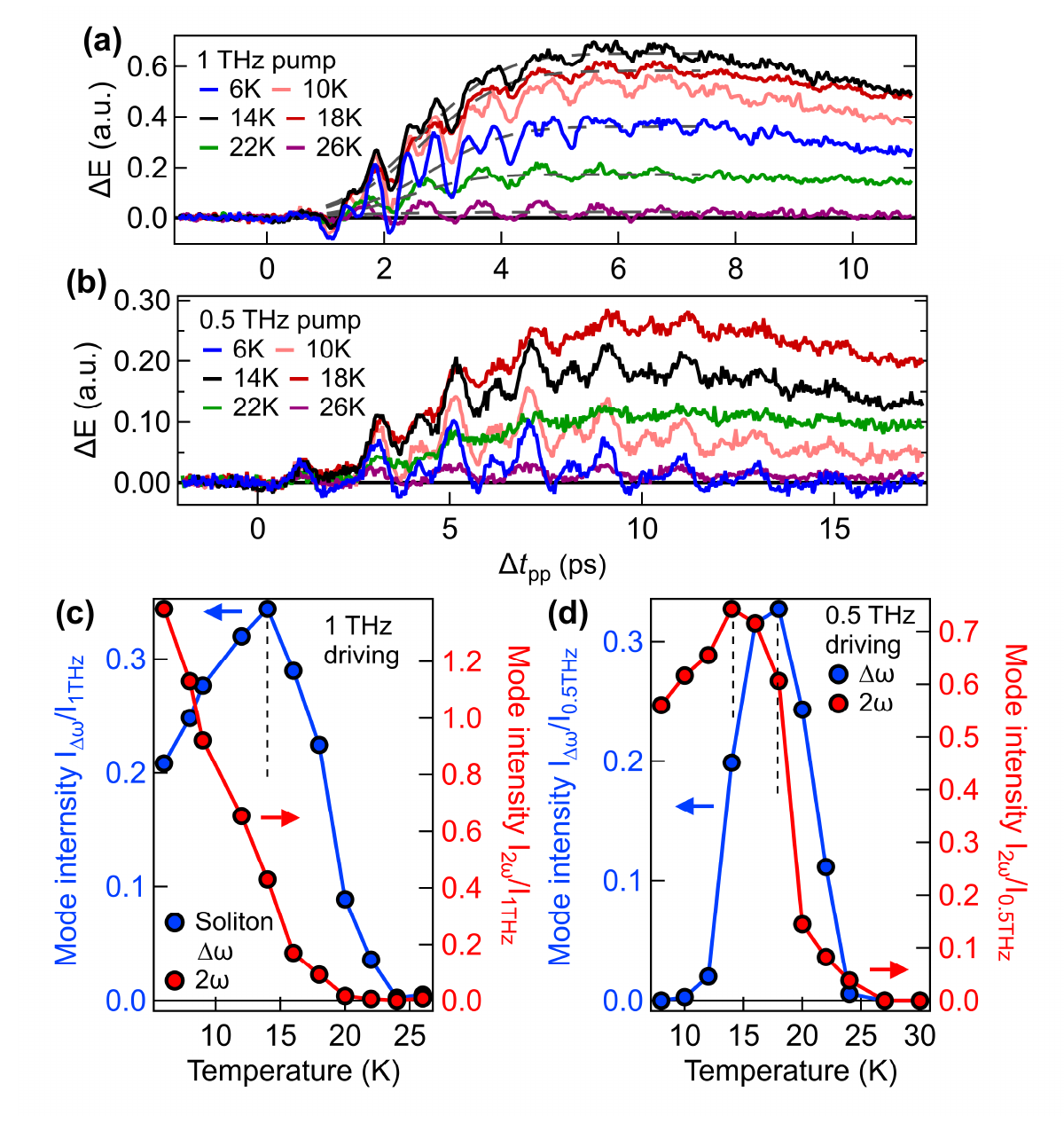}
    \caption{Temperature-dependent signatures of pseudo-spin soliton dynamics in Ba(Fe$_{0.92}$Co$_{0.08}$)$_2$As$_2$ superconductor.
    (a) Dynamics of the differential transmitted THz field $\Delta E(t_\mathrm{pp})$ measured at temperatures between 6~K and 26~K, induced by a multi-cycle narrowband pump pulse with central frequency $\omega_0 = 1$~THz and field strength $\sim 14$~kV/cm. As temperature increases toward $T_\mathrm{c}$, the overall oscillation amplitude decreases due to condensate depletion, and the oscillatory pattern transitions from multi-frequency quantum beating below $T_\mathrm{c}$ to single-frequency oscillations above $T_\mathrm{c}$.
    (b) Corresponding dynamics of $\Delta E(t_\mathrm{pp})$ at various temperatures for $\omega_0 = 0.5$~THz pump excitation. 
    (c) Temperature dependence of the normalized intensity ratios of the $2\omega_0$ peak (pseudo-spin SHG, red circles) and the $\Delta \omega$ peak (pseudo-spin soliton peak, blue circles) relative to the $\omega_0$ peak, obtained for $\omega_0 = 1$~THz pump excitation. The $2\omega_0$ peak decreases monotonically, whereas the $\Delta \omega$ peak exhibits a non-monotonic, resonant-like behavior.
    (d) Corresponding normalized intensity ratios for $\omega_0 = 0.5$~THz pump excitation. Both the $2\omega_0$ and $\Delta \omega$ peaks show non-monotonic temperature dependence, reaching maxima below $T_\mathrm{c}$. The $\Delta \omega$ peak exhibits its maximum at higher temperature than the $2\omega_0$ peak, consistent with its lower energy and thus later fulfillment of the resonance condition as the superconducting gap decreases with increasing temperature.
    }
    \label{fig2:main}  
\end{figure}

 Figure~\ref{fig2:main}(a) shows the temperature-dependent dynamics of $\Delta E(t_\mathrm{pp})$ for $T=6$~K--26~K, induced by a pump field of frequency $\omega_0=1$~THz and  strength  $\sim 14$~kV/cm. We observe three notable temperature-dependent features. First, as we approach $T_\mathrm{c}$ from below, we observe a rapid decrease in overall signal  amplitude, as expected from the depletion of the SC condensate, which diminishes above $T_\mathrm{c}$ as seen in the 26~K trace (purple). Second, of particular interest is the distinct transition in the oscillatory pattern  with increasing temperature, from a quantum beating involving multiple frequencies below $T_\mathrm{c}$ (the 6~K trace, blue) to a single-frequency, $\omega_0$ oscillation observed above $T_\mathrm{c}$ (26~K trace).
  Third, as shown by the corresponding dynamics in Fig.~\ref{fig2:main}(b) at various temperatures,   a different driving frequency, $\omega_0 = 0.5$~THz,  yields similar temperature-dependent behaviors: strong oscillatory response and multiple frequency components below $T_\mathrm{c}$, simple $\omega_0$ oscillations above $T_\mathrm{c}$.
The Fourier transform of these coherent dynamics of $\Delta E(t_\mathrm{pp})$ (Appendix~C) shows $2\omega_0$ and 
 $\omega_0$ peaks in addition to the  $\Delta \omega$ peak, same as in Fig.~\ref{fig1:main}(g), which we summarize in Figs.~\ref{fig2:main}(c) and~\ref{fig2:main}(d) for various temperatures under both 1~THz and 0.5~THz excitations. 
 Note that the data shown in Figs.~\ref{fig2:main}(c) and \ref{fig2:main}(d) 
only contains the spectral weight of the coherent oscillations in $\Delta E(t_\mathrm{pp})$ shown in Figs.~\ref{fig2:main}(a) and \ref{fig2:main}(b), and detailed in Appendix~C. 
 A slowly varying amplitude was subtracted, applied only to the temperature-dependent data without affecting the oscillatory components, to minimize overlap between the zero-frequency peak and the soliton sideband $\Delta \omega_0$ at elevated temperatures (see Sec. 3, Appendix~B).

Figure~\ref{fig2:main}(c) presents the normalized intensities of the $2\omega_0$ (red circles) and $\Delta\omega$ (blue circles) peaks, relative to the $\omega_0$ peak, as a function of temperature for an excitation frequency of $\omega_0 = 1$~THz. The $\omega_0$ peak shows a typical decrease in spectral weight with increasing temperature, approaching $T_\mathrm{c}$, and remains constant above it (Appendix~C). This trend is used to normalize the $2\omega_0$ and $\Delta\omega$ data which eliminates any temperature-dependent background due the THz transmission. Intriguingly, the $2\omega_0$ and $\Delta\omega$ peaks show non-monotonic temperature behavior, in contrast to the $\omega_0$ peak. 
We emphasize three key points. First, this  result reveals that the $2\omega_0$ peak decreases monotonically while the $\Delta \omega$ peak exhibits a non-monotonic, resonant-like dependence. These observations are consistent with the condition $\Delta \omega < 2\Delta_\mathrm{SC} < 2\omega_0$ at low temperature, such that only $\Delta \omega \approx 2\Delta_\mathrm{SC}$ can be achieved at elevated temperatures.
Second, the corresponding $\Delta \omega$ peak results for $\omega_0 = 0.5$~THz excitation are shown in Fig.~\ref{fig2:main}(d). Both display non-monotonic temperature dependence, with the $\Delta \omega$ peak reaching its maximum at a higher temperature than the $2\omega_0$ peak. This behavior is consistent with the low energy of the $\Delta \omega$ feature, which allows its resonance condition to be satisfied at higher temperatures, where the superconducting gap is smaller. 
The resonance condition under 0.5~THz excitation emerges at higher temperatures than that under 1~THz excitation, consistent with the soliton sideband picture, as $\Delta \omega$ is smaller for the 0.5~THz driving field.
Third, the $\Delta \omega$ sideband feature is strongly suppressed below 10~K under 0.5~THz excitation, consistent with the suppression of soliton collective modes due to the failure of the QP excitation and inversion condition discussed in Fig.~\ref{fig1:main}(b).
Therefore, the robust low-energy $\Delta \omega$ oscillations are sharply enhanced when the pseudo-spin soliton sideband,
$\Delta \omega$, resonates with $2\Delta_{SC}$.  We conclude that the distinct temperature and THz photon frequency dependencies of the $2\omega_0$ and $\Delta\omega$ peaks provide direct evidence for the emergence of low-energy collective excitations well below $2\Delta_{\mathrm{SC}}$, consistent with soliton collective modes and their Floquet sidebands. It is worth noting that the $2\omega_0$ and $\Delta\omega$ peaks maintain their highly distinct and non-monotonic temperature dependencies even in the raw, unnormalized data in Appendix~C.

We can clearly rule out a low-energy collective mode ($\sim$0.1 to 0.4~THz) as the origin of the $\Delta\omega$ peaks and resonance enhancement in Fig.\ref{fig2:main}, independent of the Floquet sideband mechanism. First, the 0.5~THz and 1~THz excitations produce resonant enhancements at different temperatures (dashed lines corresponding to the blue circles in Figs.~\ref{fig2:main}(c) and (d)), inconsistent with a single mode energy resonating with $2\Delta_\mathrm{SC}$. Second, 0.5~THz excitation fails to generate any $\Delta\omega$ peaks below 10~K--only laser harmonics $\omega_0$ and $2\omega_0$ are observed in Fig.~\ref{fig2:main}(d)--despite lower temperatures being more favorable for nonlinear collective mode excitation. This absence is naturally explained by the lack of strong quasiparticle excitation and Rabi flopping when $2\omega_0 < 2\Delta_\mathrm{SC}$ for 0.5~THz pumping at sufficiently low temperature, suppressing soliton formation.
Furthermore, while the $\Delta\omega$ oscillations approximately correspond to $2\Delta_\mathrm{SC}$, an additional resonance condition, $2\omega_0 = \omega_S$, also holds. Given that $\omega_S \approx 2\omega_0$ satisfies the resonance for the $2\omega_0$ peak, the two conditions become nearly indistinguishable--effectively merging into a single resonant feature, as observed in Fig.~\ref{fig2:main}(d) (red circles).

\begin{figure}[ht!]
    \centering
    \includegraphics[width=0.8\textwidth]{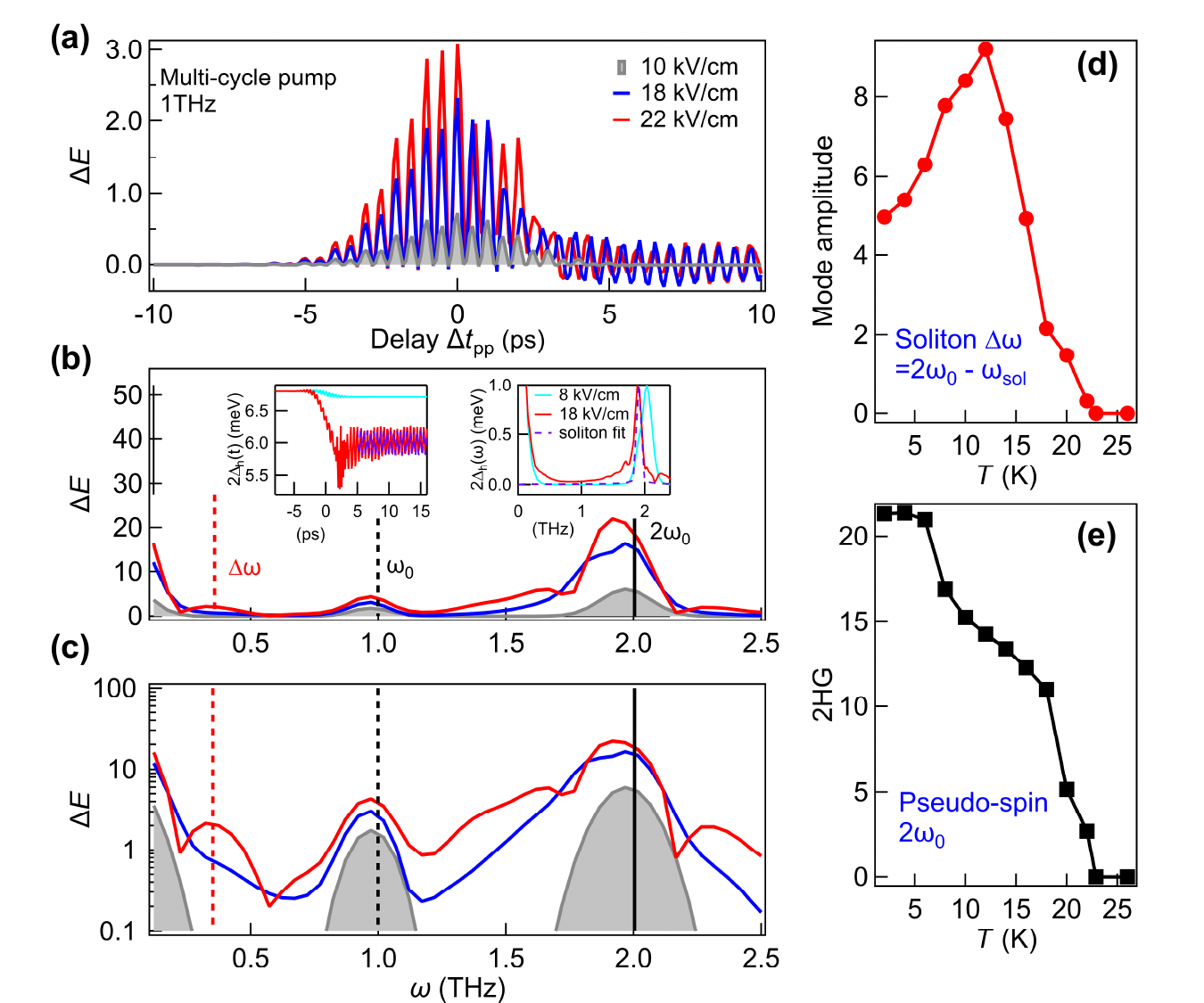}
    \caption{Simulated coherent nonlinear transmission spectra reveal soliton signatures in a multiband superconductor.
    (a) Time evolution of the simulated differential transmission signal $\Delta E$ for three different pump electric field strengths $E_0$: 10~kV/cm (gray), 18~kV/cm (blue), and 22~kV/cm (red).  The system is excited by a multi-cycle narrowband THz field with central frequency $\omega_0 = 1$~THz. At low field, oscillations are confined to the pump duration and decay rapidly; at higher fields, persistent oscillations appear after the pump, indicative of soliton dynamics.
    (b) Corresponding Fourier spectra of the signals in (a), showing peaks at $\omega_0$ (black dashed line), $2\omega_0$ (black solid line), and an emergent low-frequency sideband at $\Delta \omega$ (red dashed line) that grows with increasing field strength. Left inset: Time-dependent hole-band order parameter $\Delta_\mathrm{h}(t)$ for two different fields, revealing a transition from small quenched oscillations at $2\omega_0$ (cyan) to large, long-lived oscillations at $\omega_\mathrm{S}$ for strong fields (red). The dashed blue line indicates a fit of the post-pulse dynamics to a soliton solution using a Jacobi elliptic function, highlighting the soliton nature of the oscillations. Right inset: Corresponding Fourier spectra of $\Delta_\mathrm{h}(t)$, showing a dominant soliton peak at $\omega_\mathrm{S}$ for strong fields.  
    (c) Fourier spectra on a semi-logarithmic scale, highlighting the pronounced nonlinear growth of the $\Delta \omega$ peak compared to the fundamental and second harmonic peaks with increasing field strength.
    (d) and (e) Simulated temperature dependence of the $\Delta \omega$ peak intensity (red circles) reproducing the experimentally observed non-monotonic behavior, contrasted with (e) the monotonic decrease of the $2\omega_0$ second harmonic generation (SHG) peak (black squares) with increasing temperature. Together, these results support the interpretation of the low-frequency sideband as a hallmark of a THz-driven pseudo-spin soliton state in the multiband superconductor.}
    \label{fig3:main}  
\end{figure}

To model the experiment, we perform quantum kinetic simulations~\cite{Mootz2024} (Appendix~A) for a three-pocket SC model with strong electron--hole pocket interaction $U$ larger than the intraband pairing interaction. We chose SC order parameter values similar to those in the experimentally studied FeSC system~\cite{hybrid-higgs,Luo2023}. We directly simulate the $\Delta E$ signal measured in the experiment, which is shown in Fig.~\ref{fig3:main}(a) for three different pump fields. For weak excitation ($E_0=10$~kV/cm, shaded area), $\Delta E$ is dominated by second harmonic oscillations during the pump field driving, while the signal is zero after the pump pulse. For the two higher fields, $E_0=18$~kV/cm (blue line) and 22~kV/cm (red line), we obtain oscillations of $\Delta E$ with large amplitude during the pump pulse excitation, persisting well after the laser driving and showing only minimal decay. Figure~\ref{fig3:main}(b) shows the corresponding Fourier transformation of $\Delta E$ in Fig.~\ref{fig3:main}(a). In addition to the two conventional laser harmonic peaks centered at $\omega_0$ and $2 \omega_0$, which appear for all pump field strengths $E_0$, a third peak emerges at a low frequency $\Delta \omega$ with increasing field. The strongly nonlinear behavior of this $\Delta \omega$ peak as compared to the other two is seen more clearly in Fig.~\ref{fig3:main}(c), which plots the $\Delta E$ signal on a log-linear scale. While the  $\omega_0$ and $2\omega_0$ peaks do not show a pronounced difference between $E_0=18$ and 22~kV/cm driving,  the field dependence of the $\Delta \omega$ low-frequency peak is much stronger. 

To elucidate the origin of the $\Delta\omega$ peak, the left inset of Fig.~\ref{fig3:main}(b) shows the hole band SC order parameter $\Delta_\mathrm{h}(t)$  and illustrates the changes in its time dependence with increasing pump field. For low fields ($E_0=8$~kV/cm, cyan line), we observe a non-instantaneous small quench of this order parameter during the laser pulse, which is accompanied by second harmonic oscillations at frequency $2\omega_0$. This  behavior leads to  the two peaks in  the  spectrum $\Delta_\mathrm{h}(\omega)$ shown in the right inset of Fig.~\ref{fig3:main}(b) (cyan line), at $\omega \sim 0$ (quench) and $\omega \sim 2 \omega_0$ (second harmonic generation). For stronger fields ($E_0=18$~kV/cm, red curve),  however, this standard behavior changes. While  $\Delta_\mathrm{h}(t)$ oscillates at frequency $2\omega_0$ up to times $t \approx 0$ around the pulse maximum, persistent oscillations at a frequency $\omega_\mathrm{S}$, with  negligible damping and much stronger amplitude, develop after the pulse. To identify the origin of these oscillations, we fit their  time dependence  to that of a soliton state (Appendix~A). The order parameter  of this asymptotic non-equilibrium quantum state is time-dependent, given by $\Delta_\mathrm{S}(t)=\Delta_{+}\mathrm{dn}[\Delta_{+}(t-t_0),1-\Delta_{-}^2/\Delta_{+}^2]$ where $\mathrm{dn}$ denotes the Jacobi elliptic function, and the order parameter oscillates between $\Delta_{-}$ and $\Delta_{+}$~\cite{Yuzbashyan2008}. This soliton fit of the order parameter only applies after the pulse and yields $2\Delta_{+}=6.2$~meV and $2\Delta_{-}=5.8$~meV. The extracted soliton frequency is $\omega_\mathrm{S}=7.4$~meV, so that $\omega_\mathrm{S}< 2\omega_0$. The corresponding peak at $\omega=\omega_\mathrm{S}$ dominates the  spectrum of $\Delta_\mathrm{h}(\omega)$ in the right inset of Fig.~\ref{fig3:main}(b) (red line). Importantly, during the second part of the long multi-cycle THz pulse, $t>0$, we obtain a more complicated time dependence, with both $\omega_\mathrm{S}$ and $2\omega_0$ frequency components.
In this $t>0$ temporal regime during multi-cycle  laser pulse time-periodic driving, the soliton coherence characterized by the frequency $\omega_\mathrm{S}$ interacts  with the  QP coherence with frequency $2\omega_0$, which results in a  $\Delta E$ spectral peak at $\Delta\omega = 2\omega_0 -\omega_\mathrm{S} >0$. The calculated behavior of this sub-harmonic sideband arising during the laser driving is consistent with the experimental results presented in Fig.~\ref{fig1:main}(f).  
In particular, Figure~\ref{fig3:main}(d) reproduces the experimentally observed non-monotonic temperature dependence of the $\Delta\omega$ peak, contrasting with the monotonic decrease of the second harmonic generation  peak with increasing temperature (Fig.~\ref{fig3:main}(e)). The latter monotonic temperature dependence is expected from the non-resonant excitation condition, $2\omega_0 > 2\Delta_\mathrm{SC}$, which becomes more dramatic with growing temperature due to the decrease in the superconducting energy gap. On the other hand, the soliton state depends nonlinearly on the change of the coupling between QP and order parameter oscillations with increasing temperature, which reflects on the $\Delta \omega$ sideband peak. 

\begin{figure}[ht!]
    \centering
    \includegraphics[width=0.8\textwidth]{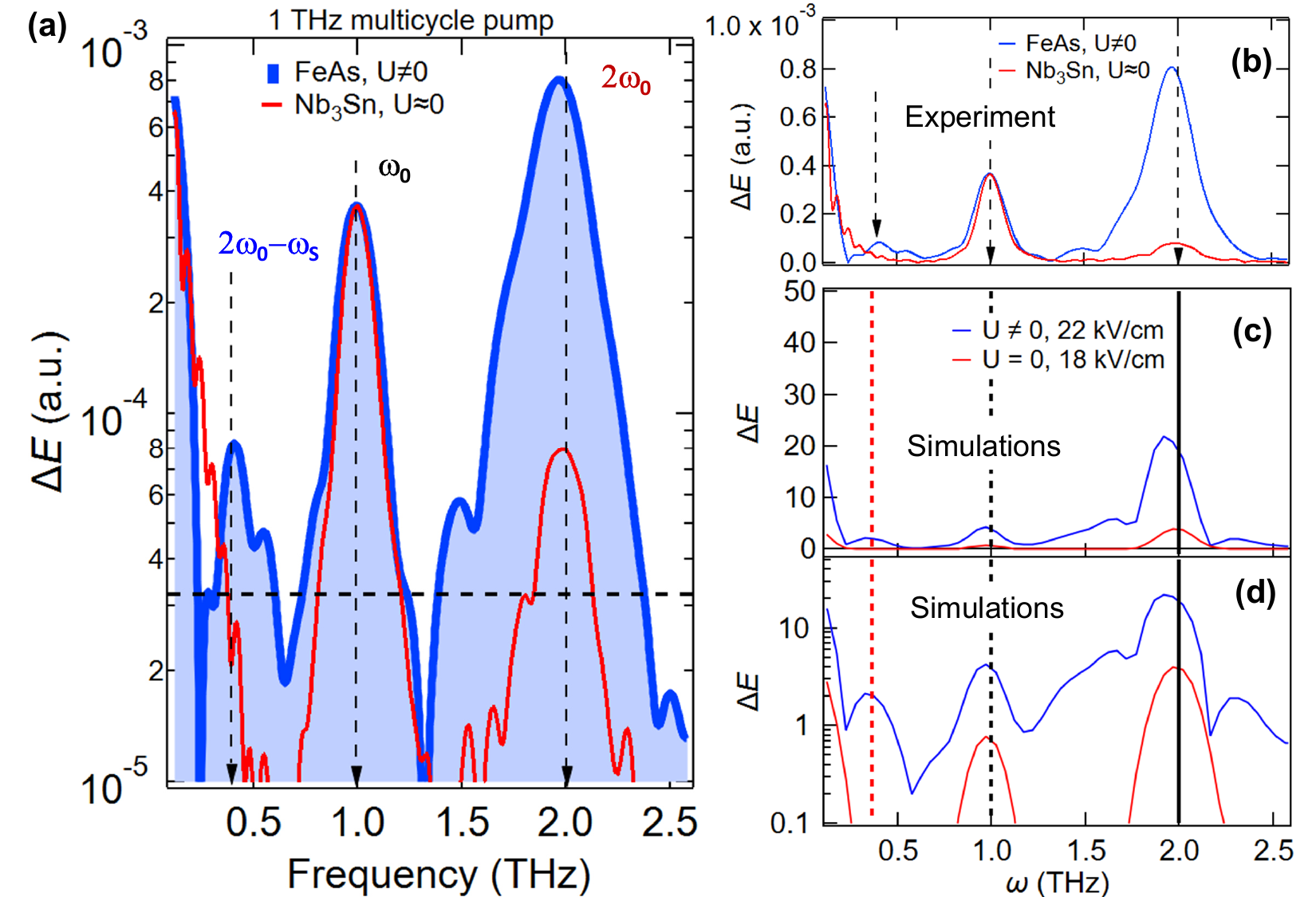}
    \caption{Comparison between multi-band and one-band superconductors under multi-cycle THz driving reveals the role of interband coupling in soliton formation.
    (a) Experimental Fourier spectrum of the coherent pseudo-spin oscillations in a one-band Nb$_3$Sn superconductor at 4.1~K, excited by a multi-cycle narrowband THz field with central frequency $\omega_0 = 1$~THz and peak electric field $\sim$~20~kV/cm (red line) and in a multi-band Ba(Fe$_{1-x}$Co$_x$)$_2$As$_2$ superconductor (blue line) at 4.1~K, both excited by a multi-cycle narrowband THz field with central frequency $\omega_0 = 1$~THz and peak electric field $\sim$~20~kV/cm. In Nb$_3$Sn, the spectrum is dominated by the fundamental $\omega_0$ peak, while the second harmonic $2\omega_0$ is suppressed and no low-frequency sideband at $\Delta \omega$ is observable. In contrast, Ba(Fe$_{1-x}$Co$_x$)$_2$As$_2$ exhibits a pronounced $2\omega_0$ peak and a clear sideband at $\Delta \omega=2\omega_0-\omega_\mathrm{S}$, consistent with soliton formation.
    (b) Same as (a) but on a linear scale to emphasize relative peak intensities. 
    (c) Simulated Fourier spectra of $\Delta E$ for the multi-band model with strong interband interaction $U \neq 0$ (blue line) and without interband interaction $U=0$ (red line), demonstrating that the $\Delta \omega$ peak (red dashed line) and enhanced nonlinear harmonics ($\omega_0$: dashed black line; $2\omega_0$: solid black line) arise from strong interband coupling.
    (d) Corresponding semi-logarithmic plot highlighting the strong nonlinear enhancement and the absence of the $\Delta \omega$ peak in the case of $U=0$. These results demonstrate that strong interband interactions in multi-band superconductors enable the formation of THz-driven pseudo-spin soliton states and associated coherent sidebands, while such states do not form in one-band systems like Nb$_3$Sn under similar excitation conditions.}
    \label{fig4:main}  
\end{figure}

Next, we  compare our findings of a Floquet pseudo-spin soliton driven state in the above multiband SC condensate  with the experimental and theoretical results of  one-band superconductors. Figure~\ref{fig4:main}(a) plots the spectrum of the coherent pseudo-spin oscillations observed in a one-band  Nb$_3$Sn SC film driven by a multi-cycle THz electric field with an amplitude of $\sim 20$~kV/cm and  central frequency of $\omega_0=1$~THz (red line). This data was taken at a low temperature of 4.1~K. For comparison, the spectrum for the multi-band Ba(Fe$_{1-x}$Co$_{x}$)$_2$As$_2$ superconductor under similar excitation conditions is also shown (blue line).
In Nb$_3$Sn, the spectrum is dominated by the fundamental $\omega_0$ peak, while the second harmonic $2\omega_0$ is strongly suppressed and no low-frequency sideband at $\Delta \omega$ is observed. In contrast, Ba(Fe$_{1-x}$Co$_{x}$)$_2$As$_2$ exhibits a pronounced $2\omega_0$ peak and a clear sideband at $\Delta \omega = 2\omega_0 - \omega_\mathrm{S}$, consistent with the formation of a pseudo-spin soliton state. Figure~\ref{fig4:main}(b) shows the same data on a linear scale, highlighting the relative peak intensities.
To motivate a possible explanation in the ideal clean system, Figs.~\ref{fig4:main}(c) and \ref{fig4:main}(d) show simulated Fourier spectra of $\Delta E$ for the multiband model with strong interband interaction ($U \neq 0$, blue line) and without interband interaction ($U=0$, red line). In these simulations, slightly different pump field strengths were used for the cases with $U = 0$ and $U \neq 0$ in order to achieve comparable initial order parameter quench amplitudes, thus ensuring a meaningful comparison of the subsequent nonlinear dynamics and spectral features.
The presence of strong interband coupling $U$ enhances the coherent nonlinear processes, leading to the emergence of the $\Delta \omega$ peak (red dashed line) and enhanced $\omega_0$ (black dashed line) and $2\omega_0$ (black solid line) harmonics. In contrast, the case with $U=0$ lacks the $\Delta \omega$ sideband and exhibits significantly weaker nonlinear harmonics.
This result for strong $U$  may explain why the soliton non-equilibrium state is difficult to realize for THz excitation of one-band superconductors, unlike for FeSCs.  

 \section{Conclusion}
 We demonstrate the realization of a laser-driven superconducting soliton state
 during multi-cycle THz driving. Unlike soliton states in atomic systems, which emerge only asymptotically after a quantum quench, our findings reveal a Floquet pseudo-spin soliton state that forms during the laser pulse in strongly coupled multi-band superconductors. This state is marked by a distinct subharmonic sideband at $\Delta\omega$, signifying collective pseudo-spin synchronization analogues to Dicke superradiance, resulting in this Floquet-like sideband during the laser pulse. The nonlinear and temperature-dependent emergence of this sideband during a multi-cycle THz excitation establishes a new quantum pathway to engineer long-lived macroscopic coherence and nonlinearities at THz speeds, with implications for quantum control, memory, and sensing.


\section*{Acknowledgments}   
\noindent {\bf Funding:} The THz experiments were supported by the National Science Foundation under Award No. 2530947. The theoretical simulations were supported by the U.S. Department of Energy, Office of Basic Energy Sciences, Division of Materials Sciences and Engineering. Ames National Laboratory is operated for the U.S. Department of Energy by Iowa State University under Contract No. DE-AC02-07CH11358.
Synthesis of pnictide thin films and characterizations at University of Wisconsin-Madison was supported by the USDOE, Office of Science, Basic Energy Sciences (BES), the Materials Sciences and Engineering (MSE) Division, under award No. DE-FG02-06ER46327.  C.B.E. acknowledges support for this research through a Vannevar Bush Faculty Fellowship (ONR N00014-20-1-2844), and the Gordon and Betty Moore Foundation’s EPiQS Initiative, Grant GBMF9065.
{\bf Author contributions:} C.V., with help of A.K., C.H. and L.L. performed experiment and analyzed raw data.
M.M., I.E.P. and J.W. developed the physical picture with discussions from all authors and M.M. performed the theoretical simulations. 
K.J.L., P.M., J.H.K., and C.B.E. grew the thin film samples and performed the structural and electrical transport characterizations.
The paper is written by J.W., M.M. and I.E.P. with help of all authors. J.W. conceived and coordinated the project.
{\bf Competing interests:} The authors declare that they have no competing interests.
{\bf Data and materials availability:} All data needed to evaluate the conclusions in the paper are present in the paper and appendices. Additional source data related to this paper is available on request from J.W.

\bibliography{ref}

\clearpage

\end{document}